\documentclass[prd,showpacs]{revtex4}
\usepackage{epsfig}
\usepackage{axodraw}
\usepackage{graphicx}

\newcommand{\gsim}{\lower.7ex\hbox{$\;\stackrel{\textstyle>}{\sim}\;$}}
\newcommand{\lsim}{\lower.7ex\hbox{$\;\stackrel{\textstyle<}{\sim}\;$}}

\begin{document}

\title{ Analysis of Leptogenesis in Supersymmetric
Triplet Seesaw Model}

\author{Eung Jin Chun$^{*\dagger}$ and Stefano Scopel$^*$}
\affiliation{$^*$ Korea Institute for Advanced Study, Seoul
130-722, Korea\\ $^\dagger$ MCTP, Physics Department, University
of Michigan, Ann Arbor, MI 48109}


\begin{abstract}
We analyze leptogenesis in a supersymmetric triplet seesaw scenario
that explains the observed neutrino masses, adopting a
phenomenological approach where
the decay branching ratios of the triplets and the amount of
CP--violation in its different decay channels are assumed as free
parameters. We find that the solutions of the relevant Boltzmann
equations lead to a rich phenomenology, in particular much more
complex compared to the non--supersymmetric case, mainly due to
the presence of an additional Higgs doublet.  Several unexpected
and counter--intuitive behaviors emerge from our analysis: the
amount of CP violation in one of the decay channels can prove to
be be irrelevant to the final lepton asymmetry, leading to
successful leptogenesis even in scenarios with a vanishing CP
violation in the leptonic sector; gauge annihilations can be the
dominant effect in the determination of the evolution of the
triplet density up to very high values of its mass, leading anyway
to a sizeable final lepton asymmetry, which is also a growing
function of the wash--out parameter $K\equiv \Gamma_d/H$, defined
as usual as the ratio between the triplet decay amplitude
$\Gamma_d$ and the Hubble constant $H$; on the other hand,
cancellations in the Boltzmann equations may lead to a vanishing
lepton asymmetry if in one of the decay channels both the
branching ratio and the amount of CP violation are suppressed, but
not vanishing. The present analysis suggests that in the
supersymmetric triplet see-saw model successful leptogenesis can be
attained in a wide range of scenarios, provided that an asymmetry
in the decaying triplets can act as a lepton--number reservoir.

\end{abstract}

\maketitle

\baselineskip 1.5em

\section{Introduction}

Leptogenesis \cite{fy} is a very appealing explanation of the
baryon asymmetry of the Universe, the more so since it can also
incorporate the origin of neutrino masses and mixings. In triplet
see-saw models, a cosmological lepton asymmetry is produced by the
decay of a heavy Higgs triplet, in presence of the Sakharov
condition of i) lepton--number violation; ii) CP violation; iii)
out--of--equilibrium decay. The same Higgs triplet is also
responsible of the dimension--five operator that induces neutrino
masses through the see-saw mechanism in the vacuum expectation
values \cite{Tss}. The lepton asymmetry is then transformed to a
baryon asymmetry through the mechanism of sphaleron conversion.

This model has already been discussed in the literature with multi
Higgs triplets \cite{Tlepto} or with additional right-handed
neutrinos \cite{hybrids}. As a general feature, it presents some
interesting differences compared to the usual see-saw mechanism
where the decaying particles producing leptogenesis are assumed to
be right--handed neutrinos. In particular, the triplet see-saw is
more predictive \cite{anna,chun03,chun05}, since Yukawa matrices
can be fixed through neutrino masses, and no unknown Majorana mass
matrices are present. An important feature of the triplet
leptogenesis is that it suffers from a strong wash--out effect due
to gauge--triplet annihilations  \cite{Hambye, chun_scopel} which
cannot be neglected, particularly, for low mass of the triplet.
Even with such a strong annihilation effect, triplets can develop
an appropriate lepton asymmetry during their thermal evolution,
and this, in the non-supersymmetric version of the model, has been
shown to lead to the possibility of a high efficiency for
leptogenesis production even for low triplet masses \cite{Hambye}.
This is an intriguing aspect that enables us to circumvent the
gravitino problem requiring the upper--bound on the reheating
temperature of the Universe, $T_{RH} \lsim 10^{8}$ GeV in
supergravity theories.

In this paper, we want to extend the discussion of the properties
of supersymmetric triplet see--saw leptogenesis to a general
phenomenological scenario, with a minimal set of theoretical
assumptions, and to compare it to the non--supersymmetric version
of the model.  In our study, we will cover the triplet mass down
to the TeV range, for which  the model has a prospect of being
tested in future colliders \cite{chun03,tev}.

The plan of the paper is as follows. In Section
\ref{section:themodel} we introduce the Higgs triplet model and
the corresponding Boltzmann equations. In Section
\ref{section:discussion} we describe our minimal assumptions and
then  discuss the produced lepton asymmetry in the
non--supersymmetric version of the model in Section
\ref{section:non_susy}, and  in the supersymmetric one in Section
\ref{section:susy}. We devote Section \ref{section:conclusions} to
our conclusions.

\section{The  model and Boltzmann equations}

\label{section:themodel}
In the supersymmetric form of the Higgs triplet model \cite{anna},
one needs to introduce vector-like pairs of
$\Delta=(\Delta^{++},\Delta^+,\Delta^0)$ and
$\Delta^c=(\Delta^{c--}, \Delta^{c-},\Delta^{c0})$ with
hypercharge $Y=1$ and $-1$, allowing for the renormalizable
superpotential as follows:
\begin{equation}
W= \lambda_L LL \Delta + \lambda_1 H_1 H_1 \Delta  + \lambda_2 H_2
H_2 \Delta^c  + M \Delta \Delta^c,
\end{equation}
where $\lambda_L LL \Delta$ contains the neutrino mass term,
$\lambda_L \nu \nu \Delta^0$. In the supersymmetric limit, the
Higgs triplet vacuum expectation value $\langle \Delta^0 \rangle=
\lambda_2 \langle H_2^0 \rangle^2/M$ gives the neutrino mass
\begin{equation}
m_\nu = 2 \lambda_L \lambda_2 {v_2^2 \over M},
\label{eq:neutrino_mass}
\end{equation}
\noindent with $v_2\equiv \langle H^0_2 \rangle$. Working in the
supersymmetric limit, we use the same notations for the bosonic
and fermionic degrees of the superfields.   A heavy particle $X$,
which can be any component of $\Delta, \Delta^c$, decays to the
leptonic final states, ${L}{L}$, as well as the Higgs final
states, ${H}_1{H}_1$ and ${H}_2{H}_2$, the
out-of-equilibrium decay of which will lead to a lepton asymmetry of
the universe.  The corresponding decay rate is $ \Gamma_X =
{|\lambda_L|^2 + |\lambda_1|^2 + |\lambda_2|^2 \over 8\pi } M$.
One of the important quantities in our analysis is $K \equiv {\Gamma_X
/ H(M)}$ which is given by
\begin{equation}
K = {|\lambda_L|^2 + |\lambda_1|^2 + |\lambda_2|^2 \over 16\pi
|\lambda_L| |\lambda_2|} {|m_\nu| M^2 \over v_2^2 H(M) } \simeq
{16 \over \sqrt{B_L B_2}}\, \left({|m_\nu| \over 0.05 \mbox{ eV}
}\right),
\label{eq:k}
\end{equation}
where $H(M)=1.66 \sqrt{g_*} M^2/m_{Pl}$ is the Hubble parameter at
the temperature $T=M$, and $B_{L,2}$ are the branching ratios of
the triplet decays to $LL$ and $H_2 H_2$, respectively.
 For the
relativistic degrees of freedom in thermal equilibrium $g_*$, we
will use the Supersymmetric Standard Model value: $g_*=228.75$.
The parameter $K$ takes the minimum value of $K_{min}=32$ for
$B_L=B_2=1/2$ and gets larger for $B_L$ or $B_2 \ll 1$. For our
discussion, we will fix $m_\nu=0.05$ eV, which corresponds to the
atmospheric neutrino mass scale.

 The resulting
lepton asymmetry of the universe is determined by the interplay of
the three asymmetries developed in the decay channels $X  \to f_i$
where $f_i=LL, H_1H_1, H_2H_2$ for $i=L,1,2$, respectively. 
Their cosmological evolutions crucially depend on the corresponding
$K$-values $K_i$ and the CP asymmetries $\epsilon_i $ which are
defined by

\begin{equation}
K_i\equiv KB_i \quad\mbox{and}\quad
\epsilon_{i} \equiv {\Gamma(X\to f_i ) - \Gamma(\bar{X}\to
\bar{f}_i ) \over \Gamma_X } \,.
\label{eq:ki_epsiloni}
\end{equation}

\noindent The above CP asymmetries follow
 the relation; $\epsilon_L + \epsilon_1 + \epsilon_2 \equiv 0$.
Note here that the model contains non-trivial CP asymmetries
$\epsilon_i$ which can be generated after integrating out
additional triplets or right-handed neutrinos
\cite{Tlepto,hybrids} or from CP phases in the supersymmetry
breaking sector \cite{softL,softT,chun_scopel}.

Before discussing how to generate leptogenesis in this model, let us
introduce for comparison its non--supersymmetric version.
The simplest realization of the triplet model corresponds to the
following Lagrangian:

\begin{equation}
{\cal L} = {\cal L}_{\rm SM} + |D_\mu \Delta|^2- M^2 |\Delta|^2+
\left (\lambda_L L L \Delta  +
M \lambda_H\, H H \Delta^*\, +\, {\rm h.c.}\right),
\label{eq:non_susy_model}
\end{equation}

\noindent with the hypercharge assignments: $Y_L=-1/2$, $Y_H=1/2$
and $Y_{\Delta}=1$. The neutrino mass term is still given by
Eq.~(\ref{eq:neutrino_mass}), with the substitutions
$v_2\rightarrow v$ and $\lambda_2\rightarrow \lambda$, and
$v\equiv \langle H^0 \rangle$).  For the relativistic degrees of
freedom in thermal equilibrium $g_*$, the corresponding Standard
Model value is $g_*=108.75$, while the total decay rate is given
by $ \Gamma_X = {|\lambda_L|^2 + |\lambda_1|^2 + |\lambda_H|^2
\over 16\pi } M$, so that the parameter $K \equiv {\Gamma_X /
H(M)}$ is given by:

\begin{equation}
K  \simeq {11.6 \over \sqrt{B_L B_H}}\, \left({|m_\nu| \over 0.05
\mbox{ eV} }\right). \label{eq:k_non_susy}
\end{equation}

\smallskip

In order to discuss how to generate a lepton asymmetry in the
supersymmetric triplet seesaw model let us first consider the
general case of a charged particle $X$ ($\bar{X}$) decaying to a
final state $j$ ($\bar{j}$) and generating tiny CP asymmetric
number densities, $n_X-n_{\bar{X}}$ and $n_j-n_{\bar{j}}$.   The
relevant Boltzmann equations in the approximation of
Maxwell--Boltzmann distributions are
\begin{eqnarray} \label{boltzmann}
 {d Y_X \over d z} &=& - z K \left[ \gamma_D (Y_X-Y_X^{eq}) +
 \gamma_A {(Y_X^2-Y_X^{eq\,2})\over Y_X^{eq}}  \right]
\nonumber\label{eq:boltzmann_X}
\\
 {d Y_x \over d z} &=& - z K \gamma_D \left[ Y_x-
 \sum_k 2 B_k {Y_X^{eq}\over Y_k^{eq}} Y_k \right]
\nonumber\label{eq:boltzmann_x}
 \\
 {d Y_j \over d z} &=& 2 z K \gamma_D\left[ \epsilon_j  (Y_X-Y_X^{eq})
 + B_j ( Y_x - 2 {Y_X^{eq} \over Y_j^{eq} } Y_j ) \right],
\label{eq:boltzmann_asym}
\end{eqnarray}
where $Y$'s are the number densities in unit of the entropy
density $s$ as defined by $Y_X\equiv n_X/s \approx n_{\bar{X}}/s$,
$Y_x \equiv (n_X-n_{\bar{X}})/s$, $Y_j\equiv (n_j-n_{\bar{j}})/s$,
and $z=M/T$. The quantities $\epsilon_i$ are defined in
Eq.~(\ref{eq:ki_epsiloni}).

The evolution of the $X$ abundance is determined by the decay and
inverse decay processes, as well as by the annihilation effect,
and are accounted for by the functions $\gamma_D$ and $\gamma_A$,
respectively.  Note that the triplets are charged under the
Standard Model gauge group and thus have a nontrivial gauge
annihilation effect which turns out to be essential in determining
the final lepton asymmetry. Moreover, as a consequence of
unitarity, the relation $2 Y_x + \sum_j Y_j\equiv 0$ holds, so
that one can drop out the equation for $Y_x$, taking the
replacement:
\begin{equation}
Y_x=-{1\over2} \sum_j Y_j,
\label{eq:sum_asym}
\end{equation}
in the last of Eqs.~(\ref{boltzmann}).
In the supersymmetric version of the model, the heavy particle $X$ can
be either of the six charged particles; $X=\Delta^{\pm\pm},
\Delta^{\pm}$ or $\Delta^{0, \bar{0}}$ for each triplets $(\Delta,
\Delta^c)$. Each of them follows the first Boltzmann equation in
Eq.~(\ref{boltzmann}) where $\gamma_D$ and $\gamma_A$ are given by
\begin{eqnarray}
\gamma_D &=& {K_1(z) \over K_2(z)} \label{eq:gamma_d}\\
\gamma_A &=& {\alpha_2^2 M \over  \pi K H(M)}
 \int^\infty_1\!\! dt\, \frac{K_1(2zt)}{K_2(z)}\, t^2 \beta(t)\, \sigma(t),
\label{eq:sigmat_int}
\end{eqnarray}
with
\begin{eqnarray}
&&\sigma(t)=(14+11 t_w^4)(3+\beta^2)+(4+ 4 t_w^2+t_w^4)\left [
16+4(-3-\beta^2 + \frac{\beta^4+3}{2\beta}\ln
\frac{1+\beta}{1-\beta})\right ]\nonumber \\ &&+4 \left
[-3+\left(4-\beta^2+\frac{(\beta^2-1)(2-\beta^2)}{\beta}\ln\frac{1+\beta}{1-\beta}\right)
\right ],\label{eq:sigmat}
\end{eqnarray}
where $t_w\equiv\tan(\theta_W)$ with $\theta_W$ the Weinberg
angle, and $\beta(t)\equiv \sqrt{1-t^{-2}}$.  The function
$\gamma_D$ is the ratio of the modified Bessel functions of the
first and second kind which as usual takes into account the decay
and inverse decay effects in the Maxwell--Boltzmann limit.  The
function $\gamma_A$ \cite{chun_scopel} accounts for the
annihilation cross-section of a triplet component $X$ summing all
the annihilation processes; $X\bar{X}^\prime \to $ Standard Model
gauge bosons/gauginos and fermions/sfermions where $X^\prime$ is
some triplet component or its fermionic partner.

The corresponding expression for $\sigma(t)$ in the
non-supersymmetric version of the model, accounting for the
annihilations of the triplets to the Standard Model gauge bosons
and fermions is given by \cite{Hambye}:

\begin{eqnarray}
&&\sigma(t)=\left (25+\frac{41}{2} t_w^4\right)\frac{\beta^2}{3}+(4+ 4 t_w^2+t_w^4)\left [
4+4(1-\beta^2 + \frac{\beta^4-1}{2\beta}\ln
\frac{1+\beta}{1-\beta})\right ]\nonumber \\ &&+4 \left
[-1+\left(2-\frac{5}{3}\beta^2+\frac{(\beta^2-1)^2}{\beta}\ln\frac{1+\beta}{1-\beta}\right)
\right ].\label{eq:sigmat_sm}
\end{eqnarray}

The r\^ole played by annihilation and decay in the determination of
the triplet density $Y_{X}$ can be understood in the following
way. When the branching ratios $B_i$ of the different decay channels are all
of the same order, inverse decays freeze out at a temperature $z_f$
determined by

\begin{equation}
K
z_f^{5/2} e^{-z_f}=1.
\label{eq:zf_K}
\end{equation}

 At that temperature the thermal averages of
the annihilation and decay rates can be compared by considering the
following ratio \cite{fry}:

\begin{equation}
\frac{<\Gamma_A>}{<\Gamma_D>}(z_f)\simeq 2 {\alpha^2\over
\alpha_X} z_f^{-3/2}e^{-z_f},
\label{eq:ratio}
\end{equation}

\noindent where $\alpha_X = K H(M)/M$. If the quantity in
Eq.~(\ref{eq:ratio}) is bigger (smaller) than 1 the triplet
freeze--out is determined by annihilation (inverse--decay).  Thus,
in this case, the annihilation effect becomes dominant for $$M
\lsim 10^{15} z_f e^{-2z_f} \mbox{ GeV},$$ and so it can be
neglected when $M\gsim 10^8$ GeV for $K=32$ and $M\gsim 1$ TeV for
$K\approx 4300$.

However, if one has $K_i\equiv B_i K<<1$ in one channel, with the
same quantity bigger than 1 in the other channels, the condition
of Eq.~(\ref{eq:zf_K}) must be modified by using $K_i$ instead of
$K$. This leads to a smaller $z_f$ and shifts to higher masses the
transition between dominance of annihilation and inverse decay in
the determination of the triplet density.

\section{Set-up for the discussion}

\label{section:discussion}

In the following, we will discuss the phenomenology of thermal
leptogenesis within the more generic version of the framework
introduced in the previous section, i.e. by considering the
branching ratios $B_i$ and the CP asymmetries $\epsilon_i$ as free
parameters, with the additional constraints $\sum B_i=1$, $\sum
\epsilon_i=0$, and $|\epsilon_i|\le 2 B_i$ (this last condition
ensures that all physical amplitudes are positive, and simply
states that the amount of CP violation cannot exceed 100\% in each
channel). This choice implies 2 free parameters in the
non--supersymmetric version of the model and 4 in the
supersymmetric one, besides the triplet mass parameter $M$. In
order to show our results, we choose to discuss, for every
particular choice of the parameters, the amount of CP violation
which is needed to provide successful leptogenesis, which we
define by the value $\bar{\epsilon}$ that the the biggest of the
$|\epsilon_i|$'s must have in order to provide $Y_L=10^{-10}$ for
the final lepton asymmetry (this amount of $Y_L$ leads to the
correct baryon asymmetry compatible to observations once
reprocessed by sphaleron interactions). Since sphaleron conversion
is suppressed at temperatures below Electro-Weak symmetry
breaking, in our calculation we stop the evolution of $Y_L$ below
$T=m_Z$, with $m_Z$ the Z--boson mass.

The quantity $\bar{\epsilon}$ is inversely proportional to the usual
efficiency factor $\eta$, defined by the relation:

\begin{equation}
Y_L=\epsilon_L \eta (Y_X+Y_{\bar{X}})|_{T>>M},
\label{eq:efficiency}
\end{equation}

\noindent where $\eta$ is determined by the amount of wash-out effect
by inverse decay and by the suppression of the number density $Y_X$ by
gauge annihilations.  One gets $\eta=1$ in the limit where the
triplets decay strongly out-of-equilibrium when $T>>M$.

By fixing $\bar{\epsilon}$ we reduce by one the number of free
parameters. Moreover, since an overall minus sign for the
$\epsilon_i$'s implies a change of sign in the final value of $Y_L$,
we discuss $Y_L$ and $\bar{\epsilon}$ in their absolute values. In this way
only the ratios among the $\epsilon_i$'s are relevant as input
parameters, and we can define the $\epsilon_i$'s in
such a way that $max(|\epsilon_i|)= 1$.

\section{The non-supersymmetric version of the model}
\label{section:non_susy}

\begin{figure}
\includegraphics[width=0.35\textwidth,angle=-90,bb = 126 31 397 766]{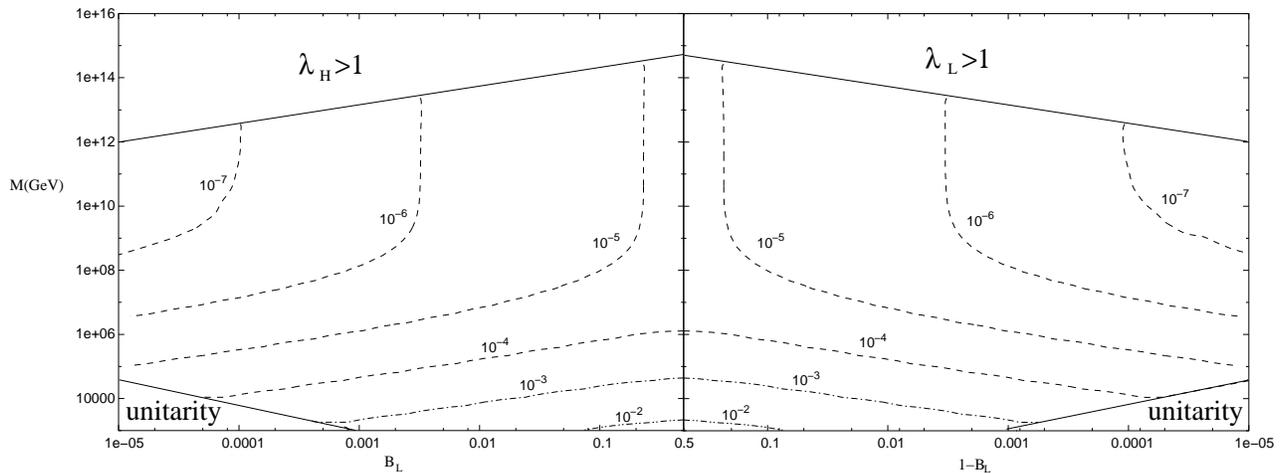}
\caption{\label{fig:contour_epsilon_bar} Contour plots of the
amount of CP violation $\bar{\epsilon}$ needed to provide the
observed baryon asymmetry in the Universe (as defined in Section
\protect\ref{section:discussion}), in the non--supersymmetric
version of the triplet see-saw model described in Section
\protect\ref{section:non_susy}, as a function of the triplet decay
branching ratio to leptons $B_L$, and of the triplet mass $M$. The
marked lower corners are excluded by the conditions
$|\epsilon_i|\le 2 B_i$, $i=L,H$, while in the upper corners one
of the Yukawa couplings becomes bigger than 1 due to
Eq.~(\protect\ref{eq:neutrino_mass}).}
\end{figure}

We start by discussing the non--supersymmetric version of the
model. In this case the triples have two two decay channels,
$X\rightarrow LL, HH$, and the process of triplet annihilations is
governed by Eq.~(\ref{eq:sigmat_sm}). In our convention CP
violations are fixed, $\epsilon_L=-\epsilon_H=1$ so there are only
2 free parameters, the triplet mass $M$ and one of the two
branching ratios, for example $B_L$, with $B_L+B_H=1$. The result
of our calculation is shown in Figure
\ref{fig:contour_epsilon_bar}, which is consistent with the result
of Ref.~\cite{Hambye}.  In our figure, the curves at constant
values of $\bar{\epsilon}$ are plotted in the $B_L$--$M$ plane (in
order to blow--up the regions where $B_L<<B_H$ or $B_L>>B_H$, the
two complementary quantities $B_L$ and $1-B_L$ are plotted in
logarithmic scale in the range 0,0.5). The marked regions in the
lower corners are excluded by the conditions $|\epsilon_i|\le 2
B_i$, $i=L,H$, while those in the upper corners are excluded
because one of the Yukawa couplings is non--perturbative due to
Eq.~(\ref{eq:neutrino_mass}). The figure is symmetric under the
exchange $B_L\rightarrow 1-B_L=B_H$, a feature that can be easily
explained by the fact that the parameter $K$ remains the same
(\ref{eq:k}) and because the identity (\ref{eq:sum_asym}) implies
$|Y_L|=|Y_H|$ at late times (when all the triplets have decayed
away).

The change of behavior of the various curves between $M\simeq
10^{8}$ and $M\simeq 10^{10}$ GeV signals a transition in the
evolution of the triplet number density between
decays/inverse--decays and annihilations. When the freeze--out
temperature of the triplets is determined by
decays/inverse--decays, $\bar{\epsilon}$ is only a function of the
branching ratios (note that the parameter $K$ does not depend on
$M$, see Eq.~(\ref{eq:k})), so curves are parallel to the vertical
axis. On the other hand, when it is the annihilation process that
determines the triplet density freeze-out, this strongly
suppresses the efficiency $\eta$ at low values of $M$ so that
higher values of $\bar{\epsilon}$ are needed in order to obtain
successful leptogenesis.

Another important feature that can be seen in the figure is given
by the fact that the highest efficiencies (lower values for
$\bar{\epsilon}$) are reached whenever $B_L<<B_H$ or $B_H<<B_L$.
As already discussed in Ref.~\cite{Hambye}, this is due to the
fact that in these cases one of the two decay channels has
$K_i<<1$ (even if, due to Eq.~(\ref{eq:k}), $K\gsim 32$), and so
is ``slow'' compared to the Hubble expansion, while the other is
``fast''. As a consequence of this, the ``slow'' channel can decay
out of equilibrium with efficiency close to 1 and develop a
sizeable asymmetry $Y_{slow}$ , while, at the same time, a
corresponding asymmetry with opposite sign $Y_x$ is left over in
the triplet density (since in this process $Y_x+Y_{slow}$ is
approximately conserved). The quantity $Y_x$ is eventually
converted into an asymmetry $Y_{fast}$ in the fast decay channel
(with $|Y_{slow}|=|Y_{fast}|$ at later times due to
Eq.~(\ref{eq:sum_asym})), when the triplets get out of kinetic
equilibrium and decay.  So, the reason why $Y_{fast}$ is not
erased by the sizeable wash-out effect is clear: due to wash-out,
triplets and anti-triplets decay practically with the same rate,
but more final particles are produced than final antiparticles
because there are more triplets than antitriplets available for
decay in the first place.  In this way an asymmetry in the triplet
density can be stored and eventually converted to a lepton
asymmetry, acting in practice as a lepton--number reservoir.  This
very simple physical picture can be significantly modified if more
than two decay channels are present, as will be illustrated in the
following sections for the supersymmetric version of the model.

\begin{figure}
\includegraphics[width=0.65\textwidth,bb =41 196 516 634 ]{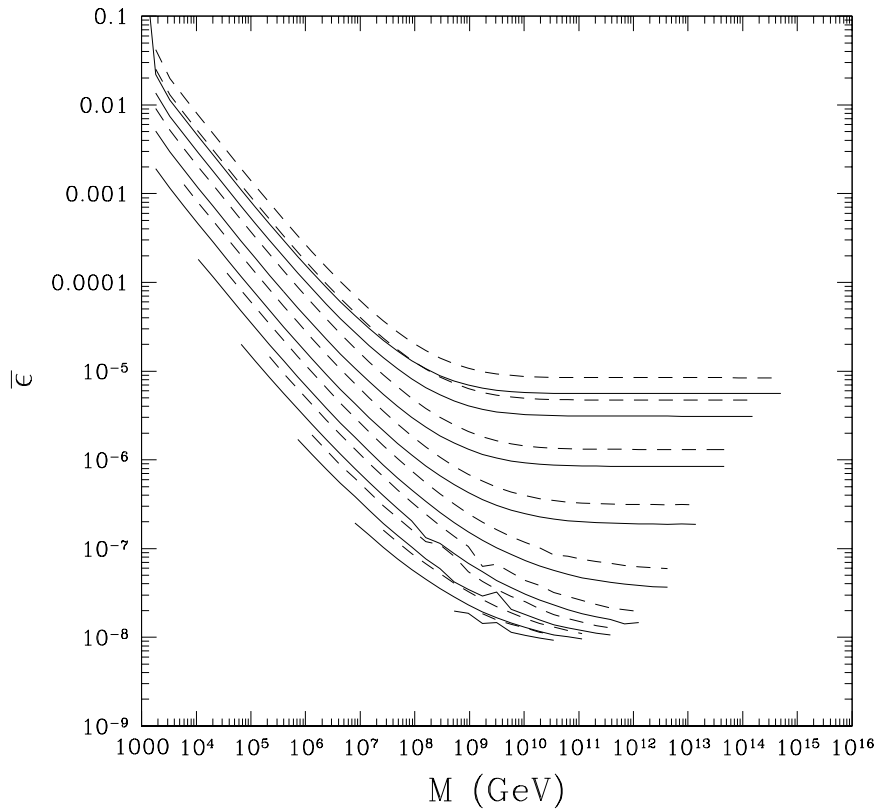}
\caption{\label{fig:SM} Amount $\bar{\epsilon}$ of CP violation
needed to provide the baryon asymmetry observed in the Universe,
as a function of the triplet mass $M$. Solid lines are calculated
in the non--supersymmetric triplet see-saw model of Section
\protect\ref{section:non_susy}, while dashed ones in the
supersymmetric model discussed in Section
\protect\ref{section:susy}, where $B_1=\epsilon_1=0$. Each curve
corresponds to a different choice of the parameter $B_L=0.5$,
$10^{-1}$, $10^{-2}$, $10^{-3}$, $10^{-4}$, $10^{-5}$, $10^{-6}$,
$10^{-7}$ and $10^{-8}$, from top to bottom. The range of each
curve is limited by the unitarity and perturbativity constraints
shown in Fig. \protect\ref{fig:contour_epsilon_bar}, as explained
in the text.}
\end{figure}

\medskip

The dependence of the quantity $\bar{\epsilon}$ as a function of
the triplet mass $M$ is discussed in Figure \ref{fig:SM}, where
the solid lines correspond to the non--supersymmetric model
discussed in this section, while the dashed ones show a
supersymmetric modification of the model discussed in the next
section. Several features of this figure are worth noticing, as
they outline quite general properties of triplet thermal
leptogenesis:

\begin{itemize}

\item As expected, $\bar{\epsilon}$ is proportional to $B_L$, and the
  higher values (lower efficiencies) are obtained when
  $B_l=B_H=1/2$.

\item The lowest value of $\bar{\epsilon}$ is reached at
  $\bar{\epsilon}\simeq 10^{-8}$. This value corresponds to the limit of
  out-of-equilibrium decay, and can be easily obtained from
  Eq.~(\ref{eq:efficiency}) by setting the efficiency $\eta=1$. In fact,
  since $Y_X|_{T>>M}\simeq 1/g_*$, where $g_*\simeq 10^{2}$ is the
  number of degrees of freedom in the Early Universe, from
  $Y_L\simeq 10^{-10}\simeq\bar{\epsilon} 10^{-2}$ one gets
  $\bar{\epsilon}\simeq 10^{-8}$. This minimal value for $\bar{\epsilon}$
  is obtained on general grounds that do not depend on
  the microphysics, so it is not expected to change
  in the modifications of the model that will be discussed in the next Sections.

\item The available range for $M$ at fixed $\bar{\epsilon}$ is bounded
from below by the unitarity constraint, and from above by the
perturbativity limit. As shown in Figure
\ref{fig:contour_epsilon_bar}, the two bounds converge at low
$B_L$ or $1-B_L$ corresponding to small values of
$\bar{\epsilon}$, and eventually meet (outside the bounds of the
figure) for $B_L=1-B_L\simeq 10^{-8}$.  That is why the range of
$M$ gets smaller for low values of $\bar{\epsilon}$, and
eventually a particular value of $M\simeq 10^{-10}$ is singled out
for which the efficiency reaches its maximum value.

\item Two different regimes for $M$ are clearly distinguishable. In
  particular, the strong loss of efficiency at lower values of $M$ is
  due to the effect of annihilations in the determination of the
  triplet freeze--out temperature. This temperature is significantly
  lowered, with a consequent suppression of the final lepton
  asymmetry, compared to the case where decays/inverse--decays
  dominate, which corresponds to the regime of higher values for
  $M$.

\item One realizes that $K$ increases but $K_L=K*B_L$ decreases from top to
bottom.  When the annihilation dominates for lower $M$, the
Boltzmann equations show that the quantity $Y_X-Y_X^{eq}$ is
determined independently of $K$ and thus the final asymmetry $Y_L$
increases with $K$.  As mentioned at the end of Section II, the
figure also shows that the dominance of inverse decay starts at
larger $M$ for smaller $K_L$.

\end{itemize}

\section{The supersymmetric version of the model}

\label{section:susy} In the supersymmetric version of the model
the particle content is enlarged, both because of the additional
supersymmetric degrees of freedom (striplets, sleptons and Higgsinos),
and from the fact that one more Higgs(+Higgsino) doublet is
included. Actually, this latter aspect will turn out to be more
relevant than the former for our discussion. In fact, barring possible
Susy--breaking effects which are expected to be suppressed for values
of $M$ above the Supersymmetry --breaking scale, triplet decay
amplitudes to particles belonging to the same supermultiplets are the
same, and can be factored out in the Boltzmann equations for
asymmetries. This implies that including supersymmetric partners in
Eqs.~(\ref{eq:boltzmann_asym}) is as simple as multiplying by 2 all
the relevant degrees of freedom, and the branching ratios $B_L$, $B_1$
and $B_2$ and CP asymmetries $\epsilon_L$, $\epsilon_1$ and
$\epsilon_2$ will refer to a sum over all the members of each
supermultiplet. As far as the triplet density $Y_X$ is concerned, the
supersymmetric version of the annihilation cross section given in
Eq.~(\ref{eq:sigmat}) must be used, where annihilations to
supersymmetric particles, as well as cohannihilations of the triplets
with their fermionic superpartners are taken into
account\cite{chun_scopel} (in our equations we assume that triplets
and striplets are degenerate in mass
). This implies a low--temperature
annihilation cross--section about a factor 8 bigger compared to the
non--supersymmetric case \cite{chun_scopel}, and a corresponding loss
of efficiency at low masses, where annihilation drives triplet
freeze--out.

\subsection{Standard Model-like case without ${\bf X\to H_1
H_1}$}
As long as only the Higgs supermultiplet $H_2$ is included
in the model (i.e., in the case with $B_1$=$\epsilon_1$=0), the
resulting phenomenology is not expected to change qualitatively
compared to the non-supersymmetric case: from the practical point
of view, the supersymmetric case with only $H_2$ corresponds just
to the non-supersymmetric one where the degrees of freedom are
multiplied by two and the annihilation cross section is about a
factor of 8 bigger (changing the degrees of freedom implies also a
slight modification of the $K$ parameter of about $\sqrt{2}$ at
fixed branching ratios). In order to show this point, in Figure
\ref{fig:SM} we show with dashed lines the result of a calculation
analogous to that shown by solid ones, where the supersymmetric
version of the model with $B_1$=$\epsilon_1$=0 is used. As can bee
seen, the two models are qualitatively quite similar, the
supersymmetric scenario implying worse efficiencies compared to
the non--supersymmetric one over the whole range of $M$. This may
be explained by the fact that in supersymmetry both the
annihilation cross section (which lowers the efficiency at low
$M$) and the $K$ parameter (which reduces it at high $M$) are
bigger compared to the non-supersymmetric case.

\subsection{Role of the third channel: ${\bf X\to H_1 H_1}$}
When non--vanishing $B_1$ and $\epsilon_1$ are considered, the number
of free parameters becomes 4 (two branching ratios and two
asymmetries) plus the triplet mass $M$. In this case, a qualitatively
different phenomenology arises compared to the previous cases.

\begin{figure}
\includegraphics[width=0.65\textwidth,bb =41 196 516 634 ]{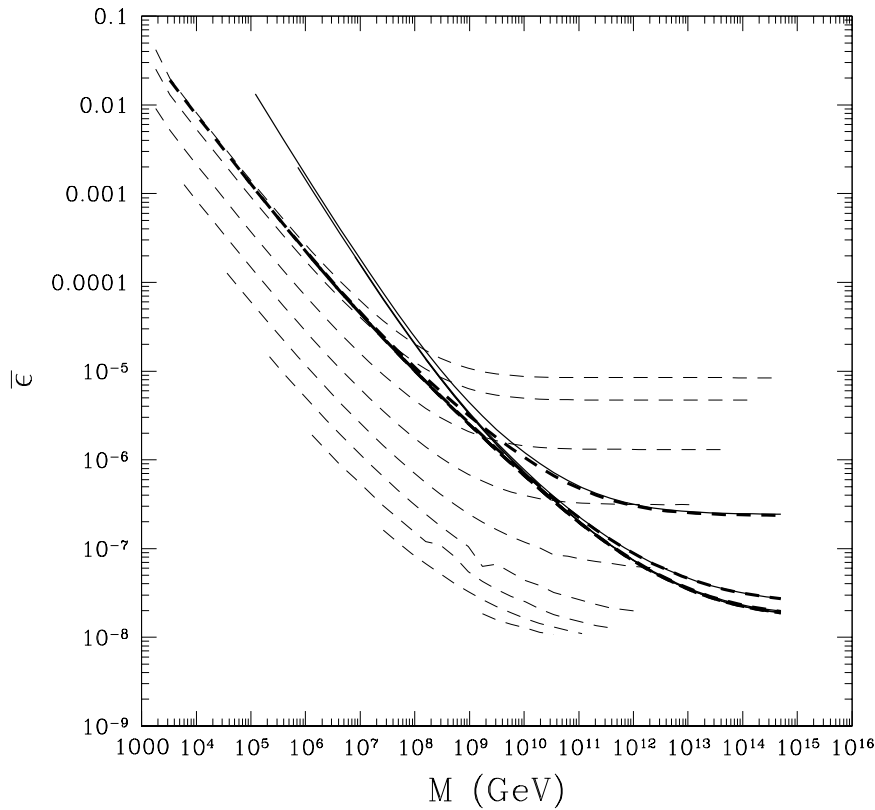}
\caption{\label{fig:high_M} Same as in Fig. \protect\ref{fig:SM}.
Thin dashed lines are calculated in the supersymmetric model discussed
in Section \protect\ref{section:susy}, where $B_1=\epsilon_1=0$,
while solid and thick--dashed lines show the same quantity for
$B_L=B_2\simeq 1/2$, $\epsilon_1=1$
and, from top to bottom, $B_1=10^{-2}$, $B_1=10^{-3}$, $B_1\le
10^{-4}$. For solid curves $\epsilon_L=0$, while 
for thick-dashed ones $\epsilon_L=-1$.}
\end{figure}

A first remarkable difference is due to the fact that $B_1$ is not
constrained by Eq.~(\ref{eq:neutrino_mass}) and can be taken
arbitrarily small even for high values of $M$. This implies that
out-of-equilibrium decay and very low values of $\bar{\epsilon}$
are expected to be reached without encountering the upper bound
on $M$ observed in the curves of Fig. \ref{fig:SM}, due to the non
perturbativity of $\lambda_L$ or $\lambda_2$.  This is shown in
Figure \ref{fig:high_M}, where $\bar{\epsilon}$ is plotted as a
function of $M$ for $B_L$=$B_2$=1/2, and for very low values of
$B_1$.

The other important feature of the model is that, as in all scenarios
with more than two decay channels, now a hierarchy in the CP violation
parameters is possible. This implies that, for instance, in some
particular channel CP violation can be suppressed compared to the
other two, or even absent. An example of this is again shown in
Fig. \ref{fig:high_M}, where the values $\epsilon_1$=1 and
$\epsilon_L=0,-1$ are assumed for each value of $B_1$. As can be seen
from Fig. \ref{fig:high_M}, even the case $\epsilon_L=0$ can provide
leptogenesis with a good efficiency. This fact at first sight might
seem quite amusing, since one could wonder how a CP--conserving decay
of the triplets to leptons may lead to any lepton asymmetry at
all. However the answer to this question is contained in the same
mechanism explained in Section \ref{section:non_susy}, where the
asymmetry in the triplet density produced by out-of-equilibrium decay
in a slow decay channel could be fully converted to an asymmetry in
the fast one even in presence of a very strong wash--out effect. As
already pointed out previously, in the fast channel CP violation
produced by triplet decays is negligible even in the case where
$\epsilon_i\ne 0$, the final asymmetry being produced only by the fact
that the number of decaying triplets is different from the number of
decaying antitriplets.  So, having $\epsilon_i=0$ or strongly
suppressed in this fast channel doesn't make any difference.

The existence of a hierarchy in the $\epsilon_i$ parameters can
however strongly affect the physical mechanism described above
whenever CP violation is suppressed in a slow channel, as can be
na\"ively expected since this is the channel that drives
leptogenesis. As a matter of fact, if the slow decay channel
has a small $\epsilon_i$ it cannot develop a sizeable asymmetry
$Y_{slow}$, while, at the same time, the corresponding asymmetry with
opposite sign $Y_x$ left over in the triplet density (due to the
approximate conservation of $Y_x+Y_{slow}$) is also suppressed,
leading so to a suppression of the asymmetry also in the other,
fast channel.

\subsection{The possibility of cancellations}
In the fast channel another important fact may arise: the two
mechanisms of asymmetry production (i.e.  direct CP violation in the
decay and asymmetry in the density of the triplets) may give rise to
effects of the same order of magnitude, and, if the sign of the
$\epsilon_i$ parameters is the same in the fast and in the slow
channels, even cancel out, leading so to a vanishing final
asymmetry. In this scenario, which implies a numerical cancellation in
the Boltzmann equations, the more populated between $X$ and $\bar{X}$
decays with the lower rate to the corresponding final state $L$ or
$\bar{L}$, in such a way that no final asymmetry is produced.

\begin{figure}
\includegraphics[width=0.65\textwidth,bb =41 196 516 634 ]{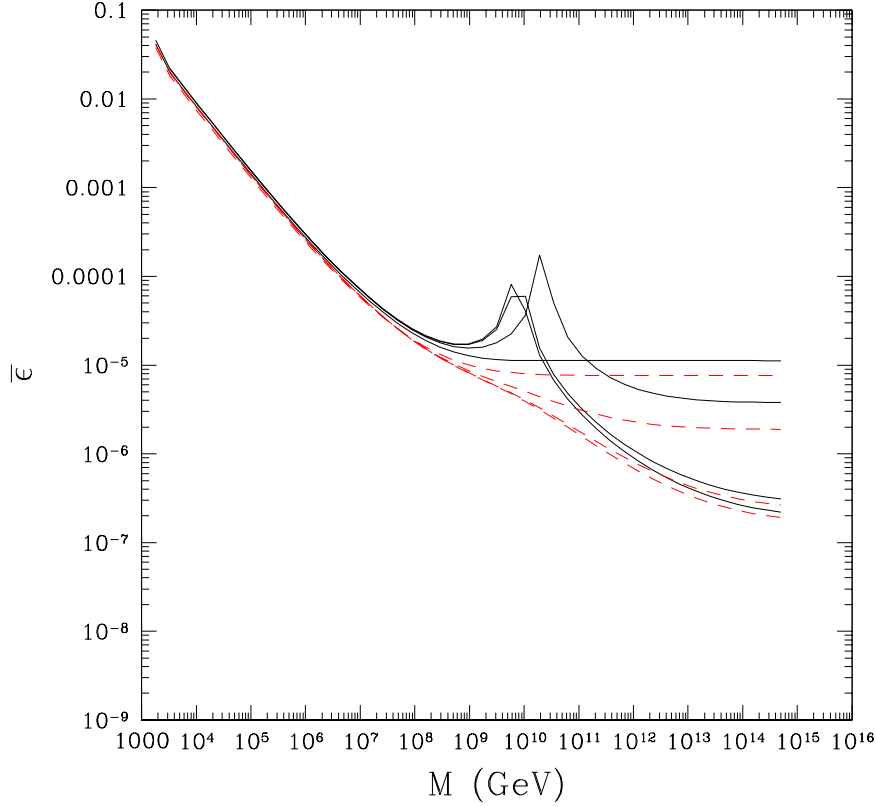}
\caption{\label{fig:peaks1} Same as in
Fig. \protect\ref{fig:high_M}, solid lines, but with $\epsilon_L=1$
and $\epsilon_1=0.1$ (solid lines) and $\epsilon_1=-0.1$ (dashed
lines). The presence of peaks in $\bar{\epsilon}$ signals a vanishing
efficiency $\eta$.}
\end{figure}

In order to show this effect, in Fig, \ref{fig:peaks1} the
parameter $\bar{\epsilon}$ is plotted as a function of $M$ for the
same choice of parameters as for the solid lines of Fig.
\ref{fig:high_M}, but assuming $\epsilon_L$=1 and $\epsilon_1=0.1$
(solid lines) and $\epsilon_1=-0.1$ (dashed lines).  As can bee
seen in the figure, now peaks arise for $\epsilon_1=0.1$
signaling a vanishing efficiency $\eta$, while are not present
for $\epsilon_1=-0.1$.  As explained before, this happens because
the $\epsilon_i$ parameter corresponding to the slowest decay
channel is suppressed compared to the other ones, and the
cancellation mechanism described in the previous paragraph may set
in when $\epsilon_1$ and $\epsilon_L$ have the same sign.

\begin{figure}
\includegraphics[width=0.65\textwidth,bb =41 196 516 634 ]{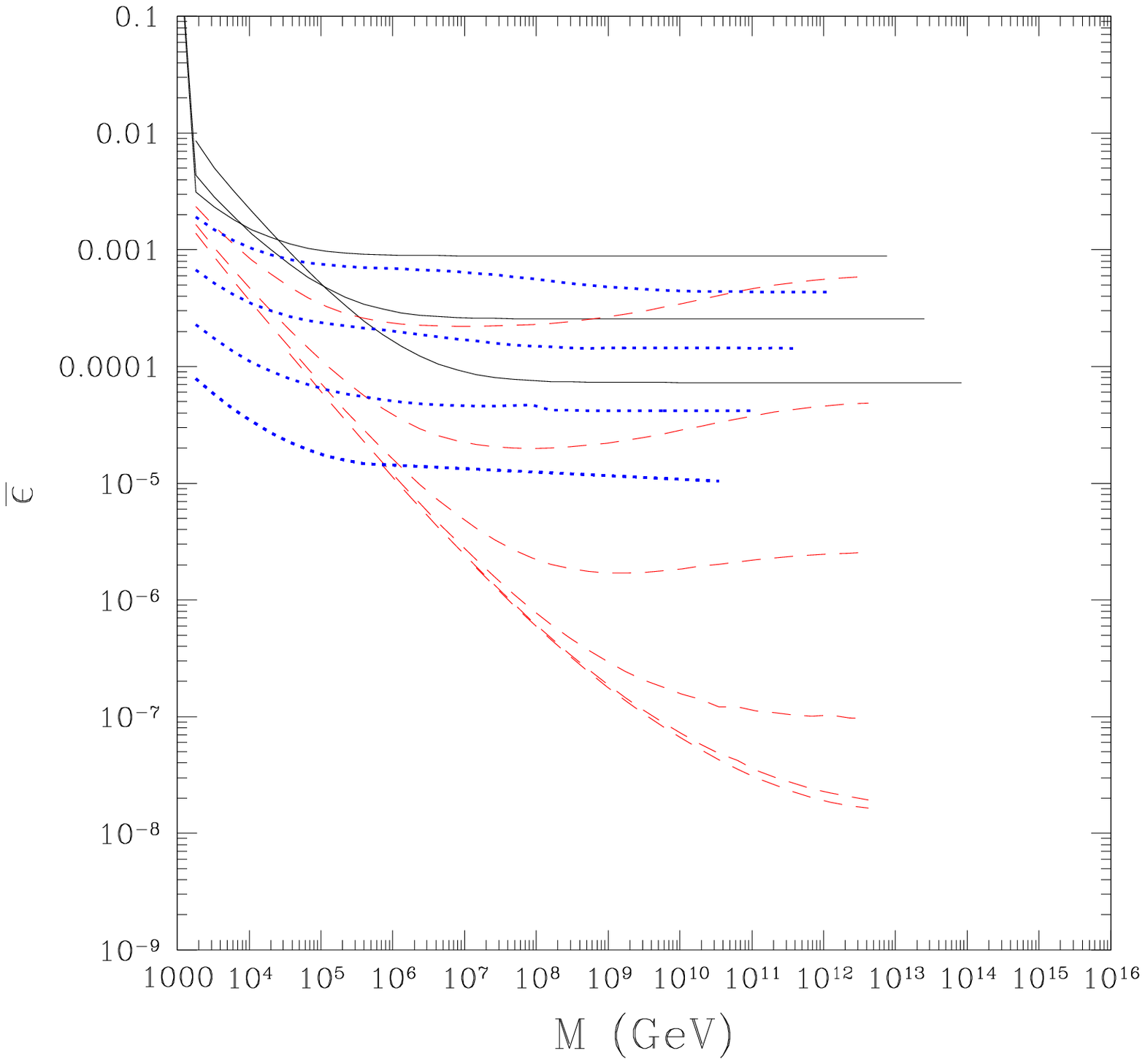}
\caption{\label{fig:epsilon2_0} Same as in Fig. \protect\ref{fig:SM},
but with $\epsilon_2$=0. From top to bottom, the solid lines correspond to the case when
$B_1$=$B_L=1/2$ and $B_2=10^{-4}$, $10^{-3}$, $10^{-2}$;
the dashed lines to $B_2=10^{-4}$ and $B_1=10^{-1}$,
$10^{-2}$, $10^{-3}$, $10^{-4}$, $10^{-5}$, $10^{-6}$;
the dotted lines to $B_2=10^{-4}$ and $B_L=10^{-1}$,
$10^{-2}$, $10^{-3}$, $10^{-4}$.}
\end{figure}

\subsection{Lepton asymmetry with vanishing ${\bf
\epsilon_2}$}
As discussed previously, when its CP--violating term is exactly
vanishing, the slow channel no longer drives leptogenesis. If its
branching ratio is also much smaller than the other two, the slow
channel may be neglected altogether. So, for instance, when
$\epsilon_1=0$, the case $B_1<<B_{L,2}$, is equivalent to taking
$B_1=0$ (so, the upper dashed curve in Fig. \ref{fig:SM}, where
$B_1=0$ and $B_L$=$B_2$=1/2, is equivalent to the case where $B_1$
is small but non--vanishing). Taking $\epsilon_{L,2}$=0 and very
small values for the corresponding $B_{L,2}$ is still equivalent
to neglecting the corresponding channel, although, due to 
Eqs.~(\ref{eq:neutrino_mass},\ref{eq:k}), a higher value for the $K$
parameter, and a possible perturbativity constraint at high $M$
are induced.

In order to show this, in Fig. \ref{fig:epsilon2_0} we show with solid
lines the case $\epsilon_2$=0, $B_2<<B_L$=$B_1\simeq$1/2. As expected,
in this scenario the efficiency is very poor, since the slow channel,
having vanishing $\epsilon_i$, cannot drive leptogenesis through
out-of-equilibriun decay. Moreover, all curves show an upper bound on
$M$ due to the perturbativity constraint, that shifts to lower $M$ for
smaller $B_2$. As already mentioned, for these curves, due to
Eq.~(\ref{eq:k}), the parameter $K$ is very high. Moreover, as
expected, $\bar{\epsilon}$ scales with $K\propto \sqrt{B_2}$ (the
efficiency is expected to scale as $1/(z_f K)$ for $K>>1$ and if the
inverse--decay effect is important \cite{fry}). It is worth noticing
now that the relative weight of the two competing effects of
annihilation and decay/inverse--decay in the determination of the
triplet freeze--out temperature depends on $K$, since the rate of
latter effect grows with $K$ while the former does not. So, the net
consequence of a big $K$ is to suppress the annihilation effect,
which, in turn, is instrumental in lowering the efficiency $\eta$ at
low values of $M$. As a consequence of this, a larger $K$ reduces the
values of $M$ where annihilation starts to dominate, as is observable
in Fig. \ref{fig:epsilon2_0}, where the change of behavior of all the
solid curves at low $M$ is shifted to the left.  At lower $M$, when
annihilation dominates in the determination of the (s)triplet density,
the efficiency grows with $K$.

If, on the other hand, in this same scenario a hierarchy between
$B_L$ and $B_1$ is assumed, the presence of another slow channel
whose corresponding CP--violation parameter $\epsilon$ is not
suppressed can in principle increase the efficiency, allowing,
eventually, to reach the values of $\bar{\epsilon}$
typical of early out-of-equilibrium decay. This effect, however, is only
possible in the case $B_1<B_L$; the alternative case $B_L<B_1$ has
a much bigger value of $K$, wich implies a better efficiency at
low mass but a worse one at higher $M$. This is shown in Fig.
\ref{fig:epsilon2_0}, where the dashed curves, which have
$\epsilon_2$=0, $B_2$=$10^{-4}$, and $B_1$=$10^{-1}$, $10^{-2}$,
$10^{-3}$, $10^{-4}$, $10^{-5}$, $10^{-6}$, from top to bottom,
can reach values as low as $\bar{\epsilon}\simeq 10^{-8}$ when
$B_1$ is sufficiently small.  On the other hand, the dotted curves
in the same figure, with $\epsilon_2$=0, $B_2$=$10^{-4}$, and
$B_L$=$10^{-1}$, $10^{-2}$, $10^{-3}$, $10^{-4}$, from top to
bottom have a worse efficiency, as expected. Note, in this last
case, the {\it inverse} proportionality between $\bar{\epsilon}$
and $K$: this is due to the fact that the relevant parameter is
$K_L\propto \sqrt{B_L/B_2}$, so that when $K$ increases $K_L$ gets
smaller.

\begin{figure}
\includegraphics[width=0.65\textwidth,bb =41 196 516 634 ]{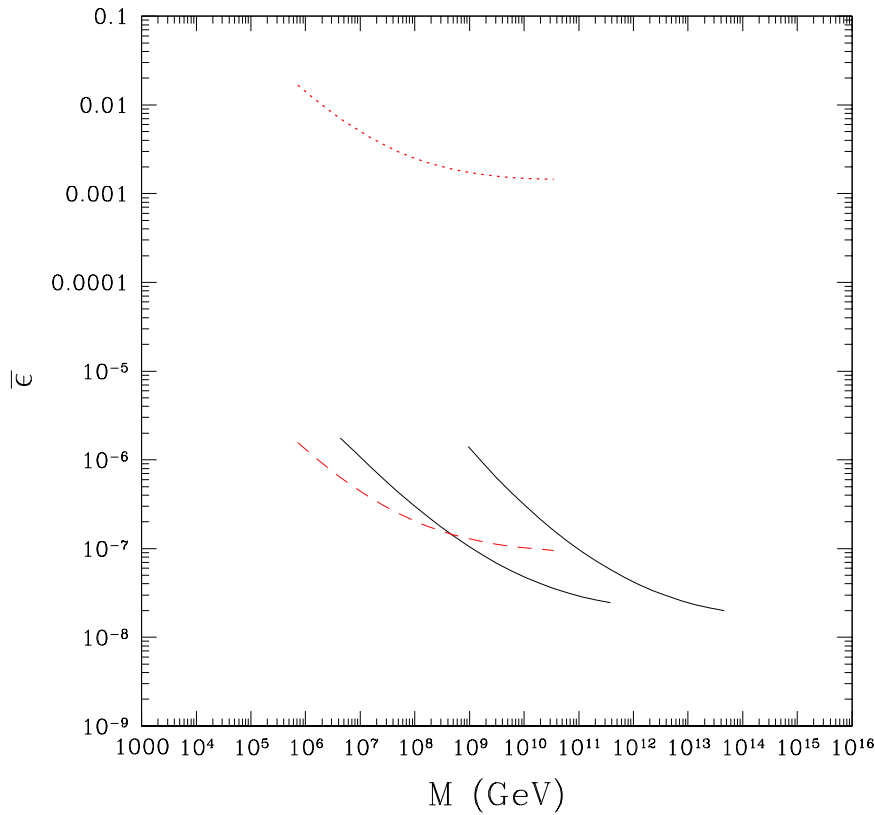}
\caption{\label{fig:epsilonl_0} Same as in Fig. \protect\ref{fig:SM},
but with $\epsilon_L$=0. The upper dotted line corresponds to
$B_L=10^{-6}$ and $B_2=10^{-2}$. For the two solid lines
$B_1=10^{-2}$, $B_2=10^{-6}$ (left) and $B_2=10^{-2}$, $B_1=10^{-6}$
(right). This last curve corresponds also to the case $B_L=10^{-2}$
$B_1=10^{-6}$. The dashed curve has $B_L=10^{-2}$ and $B_2=10^{-6}$.}
\end{figure}

\subsection{Lepton asymmetry with vanishing ${\bf
\epsilon_L}$}
As pointed out in the previous paragraph, the case $\epsilon_L$=0,
$B_L<<B_{1,2}$ is expected to give a very low efficiency for
leptogenesis. An example of this case is shown in Fig.
\ref{fig:epsilonl_0}, where the upper dotted line shows the case:
$\epsilon_L=0$, $B_L=10^{-6}$, $B_2=10^{-2}$. Eventually, taking
even smaller values of $B_L$ is equivalent to dropping $L$ from
the Boltzmann equation. On the other hand, having $\epsilon_L$=0,
$B_1<<B_{L,2}$ leads to high values of the efficiency, as already
discussed in Figure \ref{fig:high_M}. Apart from a higher value of
$K$, which implies a better efficiency at small masses but a worst
one overall, the case $\epsilon_L$=0, $B_2<<B_{L,1}$ is analogous,
since now it is the $H_2$ channel that drives leptogenesis,
decaying out-of-equilibrium with a non--suppressed $\epsilon_2$.
An example of this scenario is given by the dashed line in Fig.
\ref{fig:epsilonl_0}, where $\epsilon_L$=0, $B_2=10^{-6}$,
$B_L=10^{-2}$.

On the other hand, a qualitatively different situation is given
by: $\epsilon_L$=0, $B_{1,2}<<B_L$.  In this case there are two
slow decay channels, and the corresponding $\epsilon_{1,2}$ are
not suppressed. However, since $\epsilon_1$=-$\epsilon_2$, and,
due to Eq.~(\ref{eq:sum_asym}), at late times $Y_L=-Y_1-Y_2$, a
cancellation between $Y_1$ and $Y_2$ may occur if both quantities
reach their out-of-equilibrium value, leading so to a vanishing
$Y_L$. This implies that, in order to reach a good efficiency,
some hierarchy between $B_1$ and $B_2$ is needed, in order to have
only one slow channel. This is again shown in Fig.
\ref{fig:epsilonl_0} by the two solid lines, that correspond to:
$\epsilon_L$=0, $B_1=10^{-2}$, $B_2=10^{-6}$ (left) and
$\epsilon_L$=0, $B_1=10^{-6}$, $B_2=10^{-2}$ (right). In this case
both curves reach a good efficiency, with a less stringent
perturbativity limit for the latter due to a smaller value of $K$.
In both cases, since $K_{slow}\equiv B_{slow} K<<1$, inverse
decays in the slow channel freeze out very early, when
annihilation still dominates in the determination of the triplet
density. As a consequence of this the efficiency is a growing
function of $K$, and this explains why the solid curve on the left
($K\simeq$ 16000) is below the one on the right ($K\simeq$ 160).
Besides, in the latter case the curve would remain the same by the
exchange ($B_L \leftrightarrow B_2$), i.e.: $B_1=10^{-6}$,
$B_L=10^{-2}$. This is due do the fact that the slow channel would
remain the same, as well as the corresponding $\epsilon_i$, while
also $K$ would be unchanged.

\section{Conclusions}

\label{section:conclusions} In this paper we have analyzed the
phenomenology of  leptogenesis in the supersymmetric triplet
see--saw mechanism. Taking the branching ratios $B_i$ of the decay
rate of the triplets as free parameters, as well as the
CP--violation parameters $\epsilon_i$, with the additional
constraints $\sum_i B_1$=1 and $\sum_i \epsilon_i$=0, we have
calculated the amount $\bar{\epsilon}$ of CP violation which is
needed to provide successful leptogenesis.  In the most favourable
case of early out-of-equilibrium decay of the triplets to leptons,
this number is of order $10^{-8}$. However, it is well known that
in this scenario inverse decays and annihilations of triplets
(this last effect for lower values of $M$) contribute in general
to erase the asymmetry, basically keeping the triplets in thermal
equilibrium until late times, when $T<M$ and their number density
is exponentially suppressed. An exception to this, within the
framework of the non--supersymmetric version of the model, is
known to be the case when one branching ratio is much smaller than
the other, in such a way that one $K_i= B_i K <<$1 even if $K>>$1.
We have referred to this kind of decay channel as to a slow one,
opposed to the fast ones having $K_i>>1$.  In this case it is
sufficient that just one slow channel produces a sizeable
asymmetry, since a corresponding asymmetry is also developed in
the triplet density, which is eventually converted into an
asymmetry of the fast channel when the triplets decay. In the
supersymmetric version of the Model, this mechanism is still at
work. However, mainly because of the interplay of three decay
channels instead of two, a richer phenomenology arises:

\begin{itemize}

\item
the Yukawa coupling $\lambda_1$ of the additional Higgs doublet
can be made arbitrarily small at high triplet masses, allowing a
good efficiency also for $M\gsim 10^{12}$ GeV. In presence of only
one Higgs doublet this is not possible due to the perturbativity
bound on $\lambda_L$ implied by Eq.~(\ref{eq:neutrino_mass}).

\end{itemize}

A hierarchy among the CP--violation parameters $\epsilon_i$'s is
allowed.  Defining them in such a way that $max(|\epsilon_i|)= 1$,
and using the notation $\epsilon_{slow}$ and $\epsilon_{fast}$
for the CP--violating parameters in a slow ($K_i<<1$) and fast
($K_i>>1$) channel, respectively, this enriches the phenomenology,
because different combinations are possible:

\begin{itemize}

\item $\epsilon_{slow}=1$: The efficiency of leptogenesis reaches its
maximal value. Inverse decays in the slow channel freeze out
early, and annihilations turn out to dominate over inverse decays
in the determination of the triplet density up to quite high
values of the triplet mass $M$. In these cases the final asymmetry
is a growing function of the $K$ parameter. Moreover, the final
asymmetry is insensitive to the actual value of $\epsilon_{fast}$.
An apparently surprising example of this situation is when
$\epsilon_L=\epsilon_{fast}=0$, since in this case even a
vanishing $\epsilon_L$ can lead to efficient leptogenesis.  An
exception to this case is given by the particular situation with
$\epsilon_{slow}=1$ in {\it two} channels, namely when
$\epsilon_L<<1$ and the decay channels to $H_1$ and to $H_2$ are
both slow with $\epsilon_1=-\epsilon_2=1$. In fact, if $B_1$ and
$B_2$ are comparable, a cancellation takes place between the
asymmetries in the two channels, leading to a vanishing lepton
asymmetry, $Y_L=-Y_{H_1}-Y_{H_2}\simeq 0$.

\item $\epsilon_{slow}<1$ and  one slow channel: The final lepton
asymmetry is suppressed, as it would be na\"ively expected, since
$\epsilon_i$ is small in the only available slow channel which is
supposed to drive leptogenesis through out-of-equilibrium decays.
In this case there are two fast channels with $\epsilon_{fast}=\pm
1$, and in the channel where $\epsilon_{fast}$ has the same sign
as $\epsilon_{slow}$ this may lead to a cancellation in the
Boltzmann equations, implying a vanishing final asymmetry.
Moreover, inverse decays freeze out late in this case ($z_f\sim
\ln K >>1$), and decay is typically dominant over annihilation in
the determination of the triplet density, except for very light
values of $M$. As a consequence of this the efficiency scales as
$1/(z_f K)$ whenever $K>>1$

\item $\epsilon_{slow}<1$ and two slow channels:  Since only one
$\epsilon_i$ can be small, the other slow channel with
unsuppressed $\epsilon_i$ may drive leptogenesis with a good
efficiency. In this case, in practice the decay channel with
$\epsilon_{slow}<1$ drops out from the Boltzmann equation, and a
system with just two decay channels is recovered. However, if
$\epsilon_{slow}=\epsilon_{L,2}$, the phenomenology is different
compared to the non--supersymmetric case, because the $K$
parameter is much bigger, reducing the efficiency at high masses
and improving it at lower ones. Moreover, the unitarity limit is
more constraining at high $M$ compared to the non-supersymmetric
case.

\end{itemize}

In conclusion, the present analysis suggests that in the
supersymmetric Triplet Seesaw Model successful leptogenesis can be
attained in a wide range of scenarios, some of which appear to be non
trivial or even counter--intuitive, provided that an asymmetry in the
decaying triplets can develop at early times and be eventually
converted into a lepton asymmetry, acting in practice as a
lepton--number reservoir.


\begin{thebibliography}{99}

\def\plb#1#2#3{Phys.\ Lett.\       {\bf B#1},  #2 (#3)}
\def\npb#1#2#3{Nucl.\ Phys.\       {\bf B#1},  #2 (#3)}
\def\prd#1#2#3{Phys.\ Rev.\        {\bf D#1},  #2 (#3)}
\def\prl#1#2#3{Phys.\ Rev.\ Lett.\ {\bf #1},   #2 (#3)}
\def\rmp#1#2#3{Rev.\ Mod.\ Phys.\ {\bf #1},   #2 (#3)}
\def\mpl#1#2#3{Mod.\ Phys.\ Lett.\ {\bf A#1},  #2 (#3)}
\def\rep#1#2#3{Phys.\ Rept.\        {\bf #1},   #2 (#3)}
\def\sci#1#2#3{Science             {\bf #1},   #2 (#3)}
\def\astro#1#2#3{Astrophys.\ J.\   {\bf #1},   #2 (#3)}
\def\epj#1#2#3{Eur.\ Phys.\ J.\   {\bf C#1},   #2 (#3)}
\def\jhep#1#2#3{JHEP              {\bf #1},   #2 (#3)}
\def\ptp#1#2#3{Prog.\ Theor.\ Phys.\ {\bf #1}, #2 (#3)}
\def\ijmp#1#2#3{Int.\ J.\ Mod.\ Phys.\ {\bf A#1}, #2 (#3)}


\bibitem{fy}
M. Fukugita and T. Yanagida, \plb{174}{45}{1986}.

\bibitem{Tss} R. Barbieri, D.V. Nanopolous, G. Morchio and
F. Strocchi, Phys.\ Lett.\ {\bf B90}, 91 (1980); M. Magg and Ch.\
Wetterich, Phys.\ Lett.\  {\bf B94}, 61 (1980); J.~Schechter and
J.~W.~F.~Valle, Phys.\ Rev.~ {\bf D22} (1980) 2227; T.~P.~Cheng
and L.~F.~Li, Phys.\ Rev.\ {\bf D22}, 2860 (1980); R.N. Mohapatra
and G. Senjanovic, Phys.\ Rev.\  {\bf D23}, 165 (1981);
G.~Lazarides, Q.~Shafi and C.~Wetterich, Nucl.\ Phys.\  {\bf
B181}, 287 (1981).

\bibitem{Tlepto}
E.~Ma and U.~Sarkar, \prl{80}{5716}{1998};  T.~Hambye, E.~Ma,
U.~Sarkar, \npb{602}{23}{2001}; A. S. Joshipura, E. A. Paschos, W.
Rodejohann, \npb{611}{227}{2001}; \jhep{0108}{029}{2001}.

\bibitem{hybrids} P. O'Donnell and U.~Sarkar, \prd{49}{2118}{1994};
G.~Lazarides and Q.~Shafi, \prd{58}{071702}{1998}; E.J. Chun and
S.K. Kang, \prd{63}{097902}{2001}; T.~Hambye and G.~Senjanovic,
\plb{582}{73}{2004}; W.~Rodejohann, Phys.\ Rev.\ D {\bf 70}, 073010
(2004); P.~h.~Gu and X.~j.~Bi, Phys.\ Rev.\ D {\bf 70}, 063511 (2004);
S.~Antusch and S.~F.~King, Phys.\ Lett.\ B {\bf 597}, 199 (2004); JHEP
{\bf 0601}, 117 (2006); W. Guo, Phys.\ Rev.\ D {\bf 70}, 053009
(2004).

\bibitem{anna}
A.~Rossi, Phys.\ Rev.\  {\bf D66}, 075003 (2002).

\bibitem{chun03}
E. J. Chun, K.Y. Lee and S.C. Park, Phys.\ Lett.\  {\bf B566}, 142
(2003); A. Akeroyd and A. Aoki, \prd{72}{035011}{2005}.

\bibitem{chun05}
E. J. Chun, A. Masiero, A. Rossi, and S.K. Vempati, Phys.\ Lett.\
 {\bf B622}, 112 (2005).



\bibitem{Hambye}
  T.~Hambye, M.~Raidal and A.~Strumia,
  Phys.\ Lett.\  {\bf B632}, 667 (2006).

\bibitem{chun_scopel}
  E.~J.~Chun and S.~Scopel,
  Phys.\ Lett.\ {\bf B636} (2006) 278.



\bibitem{tev}
L. Boubekeur, T. Hambye and G. Senjanovic, \prl{93}{111601}{2004}.
A. Abada, H. Aissaoui and M. Losada, Nucl.\ Phys.\ {\bf B728}, 55
(2005); N. Sahu and U.A. Yajnik, \prd{71}{023507}{2005}; Phys.\ Lett.\
B {\bf 635}, 11 (2006); S.F. King and T. Yanagida, Prog.\ Theor.\
Phys.\ {\bf 114}, 1035 (2006); A. Pilaftsis and T.E.J. Underwood,
Phys.\ Rev.\ {\bf D72}, 113001 (2005); E.J. Chun, Phys.\ Rev.\ {\bf
D72}, 095010 (2005); S. Bray, J.S. Lee, A. Pilaftsis, Phys.\ Lett.\
{\bf B628}, 250 (2005).

\bibitem{softL}
Y. Grossman, T. Kashti, Y. Nir and E. Roulet,
\prl{91}{251801}{2003}; \jhep{0411}{080}{2004}; G. D'Ambrosio,
G.F. Giudice and M. Raidal, \plb{575}{75}{2003}.

\bibitem{softT}
G.~D'Ambrosio, T.~Hambye, A.~Hektor, M.~Raidal and A.~Rossi,
Phys.\ Lett.\  {\bf B604}, 199 (2004).



\bibitem{fry}
J.~N.~Fry, K.~A.~Olive and M.~S.~Turner,
Phys.\ Rev.\  {\bf D22}, 2977 (1980).


\end{thebibliography}
\end{document}